\documentclass[twocolumn]{aastex631}
\usepackage{CJK}
\usepackage{amsmath}
\usepackage{bm}
\usepackage{multirow,amsmath,array,booktabs}
\usepackage{color}
\usepackage{ulem}

\newcommand{\ra}{\rangle}

\newcommand{\bbm}{\begin{bmatrix}}
\newcommand{\ebm}{\end{bmatrix}}
\newcommand{\bBm}{\begin{Bmatrix}}
\newcommand{\eBm}{\end{Bmatrix}}
\newcommand{\bpm}{\begin{pmatrix}}
\newcommand{\epm}{\end{pmatrix}}

\newcommand{\nc}{\newcommand}       
\nc{\vc}[1] {\mbox{\boldmath $#1$}} 
\nc{\del}       {\partial}              
\nc{\bra}       {\langle}               
\nc{\ket}       {\rangle}               
\nc{\bras}[1]   {\langle #1|}           
\nc{\kets}[1]   {|#1\rangle}            
\nc{\mapleft}[1]{           
	\smash{\mathop{\,          %
			\hbox to 1.5cm{\rightarrowfill}\, }\limits_{#1}}}
\nc{\beq}     {\begin{eqnarray}} \nc{\eeq}    {\end{eqnarray}}
\nc{\nn}      {\\\nonumber} \nc{\vs}      {\vspace{-0.275cm}}
\nc{\fra}    {\frac{1}{2}}
\nc{\mb}        {\mathbf}

\graphicspath{{./}{figures/}}

\begin{document}
\begin{CJK*}{UTF8}{gbsn}

\title{Rotating Neutron Stars with the Relativistic {\em Ab Initio} Calculations}

\author{Xiaoying Qu}
\affiliation{School of Physics and Mechatronic Engineering, Guizhou Minzu University, Guiyang 550025, China}

\author{Sibo Wang}
\affiliation{Department of Physics and Chongqing Key Laboratory for Strongly Coupled Physics, Chongqing University, Chongqing 401331, China}

\author{Hui Tong}
\email{htong@uni-bonn.de}
\affiliation{Helmholtz-Institut f\"{u}r Strahlen- und Kernphysik and Bethe Center for Theoretical Physics, Universit\"{a}t Bonn, Bonn D-53115, Germany}

\begin{abstract}
  The equation of state (EOS) of extremely dense matter is crucial for understanding the properties of rotating neutron stars.
  Starting from the widely used realistic Bonn potentials rooted in a relativistic framework, we derive EOSs by performing the state-of-the-art relativistic Brueckner-Hartree-Fock (RBHF) calculations in the full Dirac space. 
  The self-consistent and simultaneous consideration of both positive- and negative-energy states (NESs) of the Dirac equation allows us to avoid the uncertainties present in calculations where NESs are treated using approximations.
  To manifest the impact of rotational dynamics, several structural properties of neutron stars across a wide range of rotation frequencies and up to the Keplerian limit are obtained, including the gravitational and baryonic masses, the polar and equatorial radii, and the moments of inertia.
  Our theoretical predictions align well with the latest astrophysical constraints from the observations on massive neutron stars and joint mass-radius measurements.
  The maximum mass for rotating configurations can reach up to $2.93M_{\odot}$ for Bonn A potential, while the radius of a $1.4M_\odot$ neutron star for non-rotating case can be extended to around 17 km through the constant baryonic mass sequences.
  Relations with good universalities between the Keplerian frequency and static mass as well as radius are obtained, from which the radius of the black widow PSR J0952-0607 is predicted to be less than 19.58 km.
  Furthermore, to understand how rotation deforms the equilibrium shape of a neutron star, the eccentricity is also calculated. The approximate universality between the eccentricity at the Keplerian frequency and the gravitational mass is found.
\end{abstract}


\section{Introduction}\label{Introduction}

Since the first historical detection of gravitational wave signals from the binary neutron star merger event GW170817, a transformative new avenue has opened for investigating the internal structure and properties of compact stars \citep{Abbott2017PRL,Fattoyev2018PRL,Annala2018PRL,Most2018PRL,De2018PRL,Tong:2019juo}.
Neutron stars, among the densest objects in the universe, are unique laboratories where the four fundamental forces interact mutually under extreme conditions~\citep{Lattimer2004science}.
In the cores of massive neutron stars, central densities can reach several times the nuclear saturation density. 
This extreme environment offers a rare opportunity to probe the equation of state (EOS) of highly dense matter, a critical yet poorly understood aspect of nuclear physics, particle physics, and astrophysics.

Astrophysical observations, such as precise measurements of neutron star masses, are essential for constraining the EOS and thereby enhancing our understanding of matter under extreme conditions~\citep{Watts2016RMP,Watts2019SCP}.
Three neutron stars have been measured with gravitational masses approaching twice the solar mass $2M_\odot$: PSR J1614-2230 with $M = 1.908 \pm 0.016~M_\odot$~\citep{Demorest2010Nature,Fonseca2016APJ,NANOGrav:2017wvv}, PSR J0348+0432 with $M = 2.01 \pm 0.04~M_\odot$~\citep{Antoniadis2013}, and PSR J0740+6620 with $M = 2.08 \pm 0.07~M_\odot$~\citep{Cromartie2020NA,Fonseca2021APJ}.
These measurements have ruled out those EOSs that are too soft to support a mass of $2M_\odot$~\citep{Bogdanov2019APJLa,Bogdanov2019APJLb,Miller2019,Jiang2020APJL,Raaijmakers2020APJL}.
Recent advancements by the Neutron star Interior Composition Explorer (NICER) collaboration have further refined our knowledge. 
Joint observations of the mass and radius of two millisecond pulsars, PSR~J0030+0451 \citep{Riley2019APJ, Miller2019, Vinciguerra2024APJ} and PSR J0740+6620 \citep{Riley2021APJL,Miller2021ApJ}, based on the X-ray emission from surface hot spots, have provided valuable constraints on the EOS~\citep{Raaijmakers_2019, HAN2023913_SciBull, Luo_2024_ApJ, Rutherford_2024_ApJ}.

Additionally, neutron stars can rotate at incredibly high speeds compared to other astrophysical objects due to their compactness. 
The resulting phenomena are critical for further constraints on the EOS and the nature of nuclear matter at high densities \citep{Komatsu1989MNRAS,Weber1991PLB,Salgado1994AASS,Stergioulas1995APJ,Glendenning1997PRL,Hessels2006Science,Ayvazyan2013AA,Piekarewicz2014PRC,Li2016PRD,Haensel2016EPJA,Bejger2017AA,Zacchi2017PRD}.
In the slow rotation regime, significant progress has been made with uniformly rotating neutron stars, particularly through the exploration of universal relations among quantities such as the moment of inertia, the Love number, and the quadrupole moment~\citep{Pappas2012PRL,Yagi-2013Science341.365,Yagi-2017Phys.Rep.681.1,2022-WangSB-PhysRevC.106.045804,Tong:2024egi}.
The ongoing investigation of the origins of the universal relations is expected to further enhance our understandings of the EOS.

Among the thousands of observed pulsars currently,  PSR J1748-2446ad with rotational frequency of 716~Hz~\citep{Hessels2006Science} and PSR J0952-0607 
with 707~Hz \citep{Romani_2022_APJ} are two most rapid known pulsars.
At such extreme rotational frequencies, a neutron star experiences significant centrifugal forces that cause it to deviate from spherical symmetry, resulting in an oblate shape.
This oblateness renders the Tolman-Oppenheimer-Volkoff equation~\citep{Oppenheimer1939,Tolman1939}, which assumes a static, spherically symmetric configuration, inadequate for describing rapidly rotating neutron stars.
Instead, these stars can be approximated as axisymmetric and rigidly rotating bodies within the Einstein's theory of general relativity.
Over the years, several sophisticated numerical methods for modeling (axisymmetric) rotating stellar structures have been developed \citep{Komatsu1989MNRAS,Weber1991ApJ,Weber1991PLB,Cook1992APJ,Salgado1994AASS,Stergioulas1995APJ}. These methods enable detailed exploration of the equilibrium configurations of rotating neutron stars, including their masses, radii, and moments of inertia.
Understanding these properties is crucial for interpreting pulsar observations and elucidating the underlying EOS of dense matter~\citep{Mishra2012AP,Cipolletta2015PRD,Li2016PRD,Wei2017PRD,Li2023PRC}.

As a critical element in understanding the properties and behavior of rotating neutron stars, the EOS can be theoretically derived using various nuclear many-body theories~\citep{Oertel2017,Yang2020ARNPS,BURGIO2021PPNP,Sedrakian2023PPNP,Tong:2024egi}. 
Among these, \textit{ab initio} methods with realistic nucleon-nucleon $(NN)$ interactions are particularly noteworthy because they are free from the complications of adjustable parameters. 
Relativistic \textit{ab initio} calculations are especially important, as they incorporate both kinematic and dynamic relativistic effects, which are significant at the high densities characteristic of neutron star cores. 
The relativistic Brueckner-Hartree-Fock (RBHF) theory has been instrumental in addressing the challenge of \textit{ab initio} methods within a relativistic framework and has received considerable attention~\citep{Tong:2018qwx, Tong2022ApJ, Tong:2022tlt, WANG-SB2021_PRC103-054319, WangS2022PRC-L021305, Farrell2024ApJ, CCWang-2024-PhysRevC.109.034002, Zou2024PLB, QinPP-2024-PhysRevC.109.064603, WangSB2024NPR}. 
Compared to its non-relativistic counterpart, RBHF theory is distinguished for its ability to reproduce the empirical saturation properties of symmetric nuclear matter without the need for additional three-body forces~\citep{Brockmann1990}.

A crucial aspect of RBHF calculations is the self-consistent determination of the single-particle potential for nucleons within the nuclear medium, which is achieved through the effective $G$ matrix. 
Recent advancements in RBHF theory have enhanced its capability by incorporating both positive- and negative-energy states (NESs) within the Dirac space~\citep{WANG-SB2021_PRC103-054319}. 
These improvements facilitate a detailed decomposition of the matrix elements in the full Dirac space, providing unique insights into the Lorentz structure and momentum dependence of the single-particle potential. 
Notably, this approach has resolved a long-standing controversy concerning the isospin dependence of the single-particle potential~\citep{WangS2022PRC-L021305}, clarifying discrepancies that arose from earlier methods that incorporate NESs with momentum-independent approximation (mom.-ind. app.) and projection techniques. 
These improvements have also lead to a more comprehensive and self-consistent calculation of the EOS for both symmetric and asymmetric nuclear matter, and static neutron star properties~\citep{WANG-SB2021_PRC103-054319,WangS2022PRC-L021305,2022-WangSB-PhysRevC.106.045804,Tong2022ApJ,2023-QuXY-SciChina}.

In this work, it is timely to study the properties of rotating neutron stars with the latest RBHF theory in the full Dirac space.
This paper is arranged as follows. The theoretical framework of the RBHF theory in the full Dirac space, EOS for neutron star, and the rotating neutron star are briefly introduced in Section~\ref{Theoreticalframework}. 
In Section~\ref{Resultsdiscussions}, the properties of rotating neutron star and related discussions are presented. The summary is given in Section~\ref{summary}.

\section{Theoretical framework}\label{Theoreticalframework}

\subsection{Relativistic Brueckner-Hartree-Fock Theory}

The RBHF theory describes the motion of a single nucleon in nuclear matter using the Dirac equation
\begin{equation}\label{DiracEquation}
  \left\{ \bm{\alpha}\cdot\bm{p}+\beta \left[M+\mathcal{U} (\bm{p})\right] \right\} u (\bm{p},s)
  = E_{\bm{p}}u(\bm{p},s),
\end{equation}
where $\bm{\alpha}$ and $\beta$ are the Dirac matrices, $M$ is the nucleon mass, $\bm{p}$ is the momentum, and $E_{\bm{p}}$ is the single-particle energy. 
The spin index is denoted by $s$. The isospin index is omitted for simplicity in this context.
According to symmetry analysis~\citep{SerotWalecka1986}, the single-particle potential $\mathcal{U} $ in the infinite uniform nuclear matter can be expressed as
\begin{equation}\label{SPP}
  \mathcal{U} (\bm{p}) = U_{S}(p)+ \gamma^0U_{0}(p) + \bm{\gamma\cdot\hat{p}}U_{V}(p),
\end{equation}
where $U_{S}$ is the scalar component of the single-particle potential, $U_{0}$ represents the timelike part, and $U_{V}$ denotes the spacelike part of the vector potential. 
Here, $\hat{\bm{p}}=\bm{p}/|\bm{p}|$ is the unit vector in the direction of $\bm{p}$. 
By introducing the effective mass $M^*_{\bm{p}}= M+U_{S}(p)$, the effective momentum $\bm{p}^* = \bm{p}+\hat{\bm{p}}U_{V}(p)$, and the effective energy $E^*_{\bm{p}}= E_{\bm{p}}-U_{0}(p)$, the analytical solution of Equation~\eqref{DiracEquation} provides the in-medium Dirac spinor
\begin{equation}\label{DiracSpinor}
  u (\bm{p},s) =\ \sqrt{\frac{E_{\bm{p}}^*+M_{\bm{p}}^*}{2M_{\bm{p}}^*}}
  		\bbm 1 \\ \frac{\bm{\sigma}\cdot\bm{p}^* }{E_{\bm{p}}^*+M_{\bm{p}}^*}\ebm \chi_s,
\end{equation}
where $\chi_s$ is the spin wave function.

The mean field in which a nucleon moves is characterized by the single-particle potentials and is generated from the effective $NN$ interactions in the nuclear medium. 
In the RBHF theory, this effective $NN$ interaction is represented by the $G$ matrix, which results from an infinite summation of ladder diagrams of the realistic $NN$ interaction. 
In practice, the $G$ matrix is obtained by solving the in-medium $NN$ scattering equation
\begin{widetext}
\begin{equation}\label{eq:ThomEqu}
  \begin{split}
  G(\bm{q}',\bm{q}|\bm{P},W)
  = V(\bm{q}',\bm{q}|\bm{P})
  + \int \frac{d^3k}{(2\pi)^3}
  V(\bm{q}',\bm{k}|\bm{P})
    \frac{Q(\bm{k},\bm{P})}{W-E_{\bm{P}+\bm{k}}-E_{\bm{P}-\bm{k}}}  G(\bm{k},\bm{q}|\bm{P},W).
  \end{split}
\end{equation}
\end{widetext}
Here $\bm{P}=({\bm k}_1+{\bm k}_2)/2$ denotes the half of the total momentum, and $\bm{k}=({\bm k}_1-{\bm k}_2)/2$ is the relative momentum of two interacting nucleons with momenta ${\bm k}_1$ and ${\bm k}_2$. The Pauli operator $Q$ prevents the intermediate states being occupied states below the Fermi surface, and the starting energy $W$ corresponds to the total single-particle energies in the initial states.

The solution of the $G$ matrix necessitates the realistic $NN$ interaction $V$ as input. 
In this work, we employ three parameterizations of the Bonn potentials, i.e., Bonn A, B, and C~\citep{Machleidt1989}, as our previous studies on non-rotating neutron stars~\citep{WangS2022PRC-L021305,2022-WangSB-PhysRevC.106.045804,Tong2022ApJ,2023-QuXY-SciChina}.
It is important to note that the $NN$ interaction $V$ in the nuclear matter is constructed in terms of effective Dirac spinors~\eqref{DiracSpinor}, thereby incorporating density dependence in a self-consistent way. 
This is in contrast to its non-relativistic counterpart where the two-body interaction is independent on densities.
From the $G$ matrix, extracting the three components of single-particle potential operator, $U_S, U_0, U_V$, is a crucial step in RBHF calculations. 
In this work, we adopt the method rooted in the full Dirac space, which takes into account both the positive- and negative-energy states of in-medium Dirac spinors~\citep{WANG-SB2021_PRC103-054319,WangS2022PRC-L021305}. This approach involves computing the $V$ matrix and $G$ matrix between positive- and negative-energy states. 
By doing so, we can construct the matrix elements of the single-particle potential operator $\mathcal{U}$ in the full Dirac space, thereby determine uniquely the Lorentz structure of $\mathcal{U}$.

After the solution of $G$ matrix and the calculation of single-particle potentials converge, the binding energy per nucleon in nuclear matter can be calculated using
\begin{widetext}
\begin{equation}\label{E/A}
  \begin{split}
  E/A
  =&\ \frac{1}{\rho} \sum_{s} \int^{k_F}_0 \frac{d^3p}{(2\pi)^3} \frac{M^*_{\bm{p}}}{E^*_{\bm{p}}}
  \langle \bar{u} (\bm{p},s)| \bm{\gamma}\cdot\bm{p} + M |u (\bm{p},s)\ra - M \\
    &\ + \frac{1}{2\rho} \sum_{s,s'} \int^{k_F}_0 \frac{d^3p}{(2\pi)^3} \int^{k'_F}_0
     	  \frac{d^3p'}{(2\pi)^3} \frac{M^*_{\bm{p}}}{E^*_{\bm{p}}}\frac{M^*_{\bm{p}'}}{E^*_{\bm{p}'}}
  	\langle \bar{u} (\bm{p},s) \bar{u}(\bm{p}',s') |\bar{G}(W)| u (\bm{p},s) u(\bm{p}',s') \ra.
  \end{split}
\end{equation}
\end{widetext}
In this expression, $\bar{G}$ denotes the antisymmetrized $G$ matrix, while $\rho$ and $k_F$ denote the total density of nucleons and the Fermi momentum, respectively.

\subsection{Equation of State for Neutron Star}

In this work, the neutron star is described as a $\beta$-stable nuclear matter system composed of nucleons and leptons (mainly electrons and muons).
The equilibrium conditions for the chemical potentials of nucleons and leptons are given by
\begin{equation}\label{chemequilb}
  \mu_p = \mu_n - \mu_e, \quad \mu_\mu=\mu_e,
\end{equation}
where $\mu_e$, $\mu_\mu$, $\mu_p$, and $\mu_n$ denote the chemical potentials of electrons, muons, protons, and neutrons, respectively.
Charge neutrality is maintained as
\begin{equation}\label{chargneut}
  \rho_p = \rho_e + \rho_\mu,
\end{equation}
where $\rho_p, \rho_e$, and $\rho_\mu$ are the number densities of protons, electrons, and muons, respectively. 
The energy density of the $\beta$-stable nuclear matter is then obtained as
\begin{equation}\label{nm-ene}
  \varepsilon = \rho \left[E(\rho,\alpha)/A + Y_p M_p + (1-Y_p)M_n \right]+\varepsilon_e+\varepsilon_\mu,
\end{equation}
where $Y_i = \rho_i/\rho~(i = e,~\mu,~p,~n)$ are the equilibrium particle fractions, and $\alpha=(\rho_n-\rho_p)/\rho$ is the asymmetry parameter.
The chemical potential for each particle $i$ is given by
\begin{equation}\label{se-chempot}
  \mu_i = \frac{\partial \varepsilon/\rho}{\partial Y_i}.
\end{equation}
For a given density $\rho$, the particle fractions $Y_i$ are determined by solving the equilibrium conditions and charge neutrality, allowing the calculation of the energy density $\varepsilon$ using Equation \eqref{nm-ene}. 
The pressure $P$ of the $\beta$-stable nuclear matter is then derived from
\begin{equation}\label{nm-pre}
  P=-\frac{\partial (\varepsilon/\rho)}{\partial (1/\rho)}=\rho\frac{\partial \varepsilon}{\partial \rho}-\varepsilon.
\end{equation}
This yields the EOS of $\beta$-stable nuclear matter in the form of $P(\varepsilon)$, which serves as the input for determining neutron star properties, together with the EOS from Baym-Pethick-Sutherland~\citep{Baym1971-BPS} and the Baym-Bethe-Pethick~\citep{Baym1971-BBP} model for crust.

\subsection{Rotating Neutron Star}

To solve for rotating and axisymmetric neutron star configurations within general relativity, we consider the stellar matter as a perfect fluid described by the energy-momentum tensor given by

\begin{equation}
  T^{\mu\nu}=(\varepsilon + P)u^{\mu}u^{\nu}-g^{\mu\nu}P,
\end{equation}
where $\varepsilon$, $P$, and $u^{\mu}$ are the energy density, pressure, and fluid's four-velocity, respectively.
We address the Einstein field equations for an axisymmetric and stationary space-time with the metric
\begin{equation}
  \begin{split}
  	ds^2=&\ -e^{\gamma+\rho}dt^2+e^{2\alpha}(dr^2+r^2d\theta^2)\\
  		&\ +e^{\gamma-\rho}r^2\sin^2\theta(d\phi-\omega dt)^2,
  \end{split}
\end{equation}
where the metric potentials $\gamma, \rho, \alpha$, and $\omega$ are functions of the radial coordinates $r$ and the polar angle $\theta$. For numerical integration of the equilibrium equations, we utilize the RNS code~\citep{Stergioulas1995APJ, Paschalidis2017LR} to compute equilibrium configurations of rotating neutron stars, including their masses, radii, and moments of inertia, given a specified central energy density.

\section{Results and discussions}\label{Resultsdiscussions}

\begin{figure*}[htbp]
  \centering
  \includegraphics[width=0.45\textwidth]{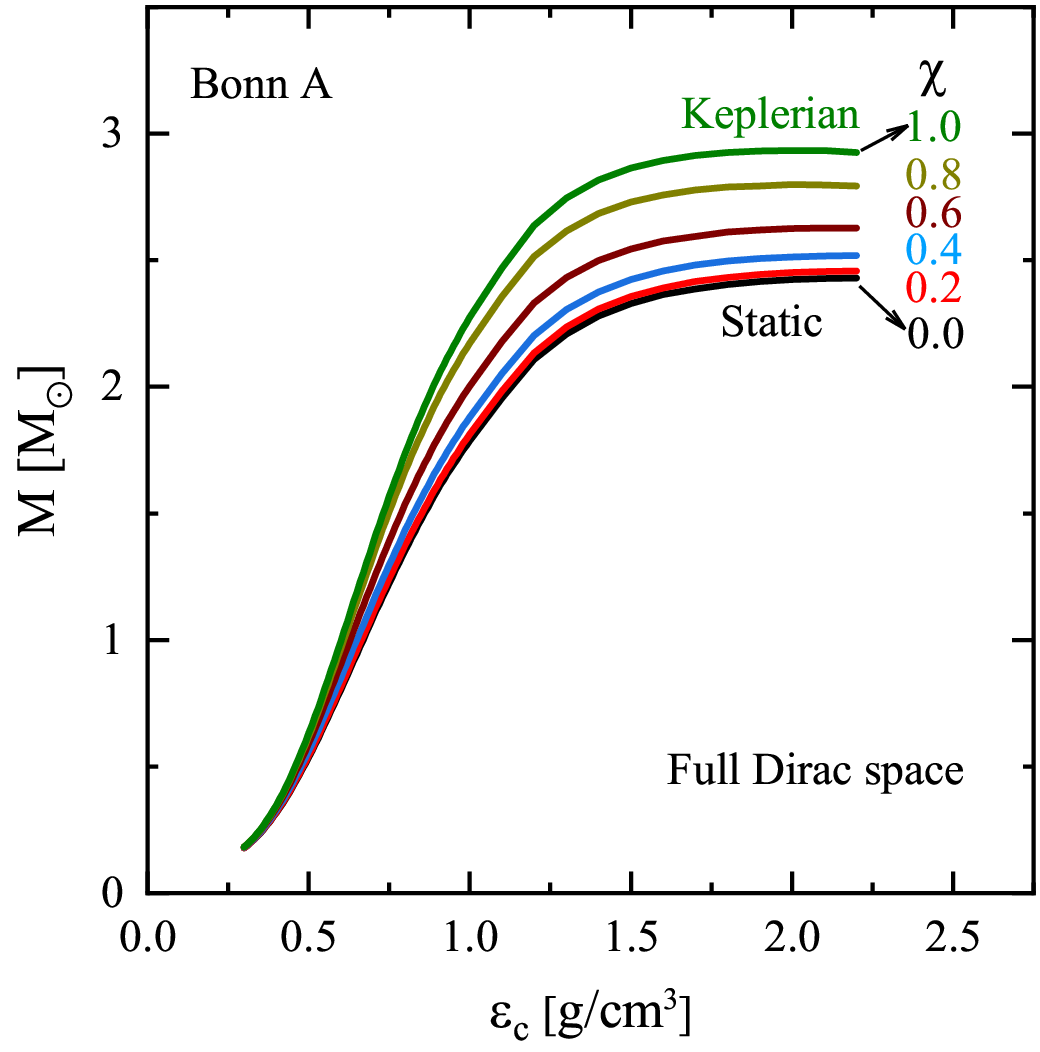}
  \hspace{0.5cm}
  \includegraphics[width=0.45\textwidth]{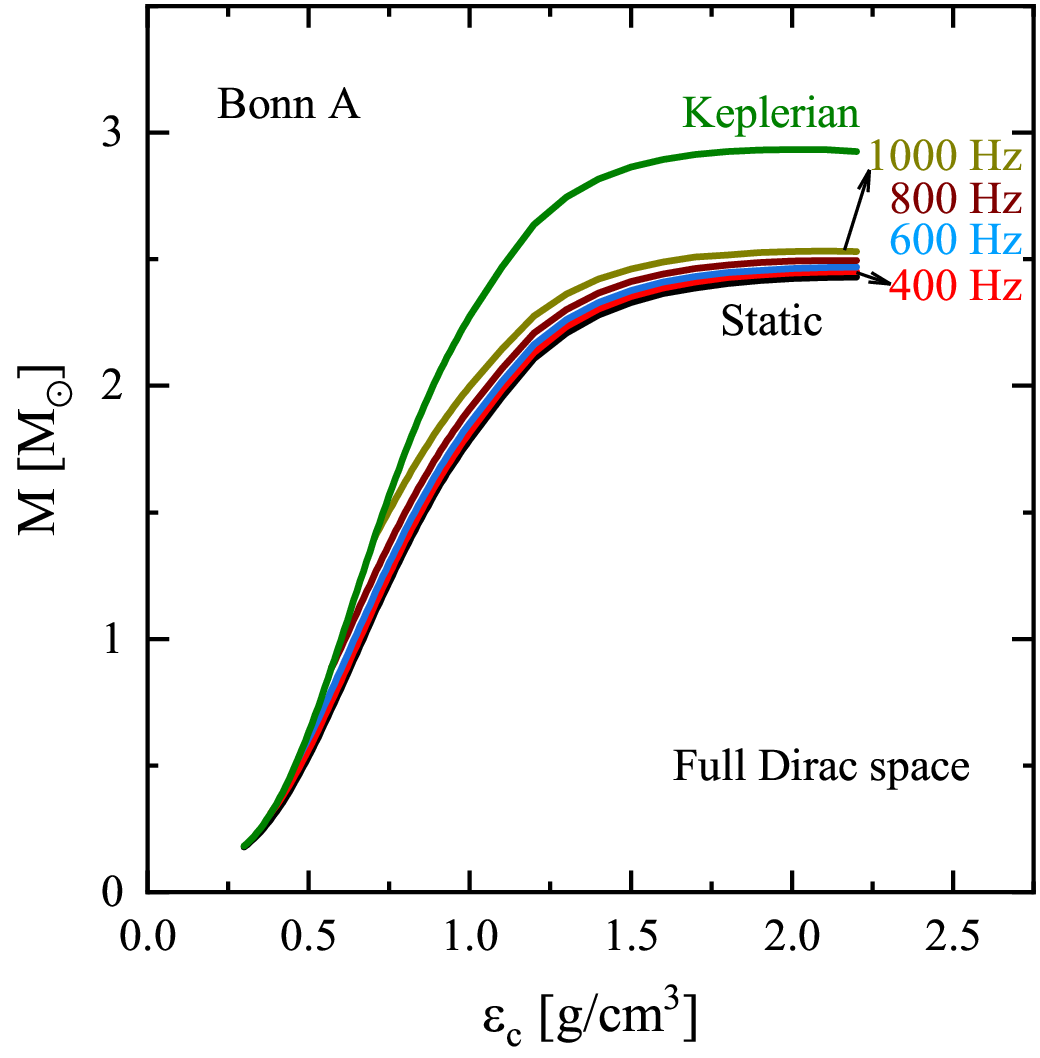}
  \caption{
  Gravitational mass $M$ as a function of central energy density $\varepsilon_c$. Six cases are shown for (left panel) constant spin ratios $\chi \equiv f/f_K = 0$ (static), 0.2, 0.4, 0.6, 0.8, and 1.0 (Keplerian sequence) and (right panel) constant spin frequencies $f = 0, 400, 600, 800, 1000$ Hz and $f=f_K$, where $f_K$ denotes the Keplerian frequency. These calculations are based on the EOSs from RBHF calculations in the full Dirac space with the Bonn A potential.
  }
  \label{fig1}
\end{figure*}

The impact of rotation on stellar structures significantly influences the maximum mass a star can sustain at a given central energy density.
Figure \ref{fig1} depicts how the gravitational mass varies with central energy density for both static and rotating neutron stars, using the EOS derived from RBHF calculations in the full Dirac space with the Bonn A potential.
The left panel displays cases with constant spin ratios $ \chi \equiv f/f_K = 0.0, 0.2, 0.4, 0.6, 0.8, 1.0$, where the Keplerian frequency $f_K$ is the threshold at which the star's mass reaches its maximum due to the balance between centrifugal and gravitational forces, beyond which the star cannot remain stable. In parallel, the right panel highlights scenarios with constant spin frequencies $f = 0, 400, 600, 800, 1000$ Hz, and $f=f_K$.

For a given central energy density, the presence of rotation enables a neutron star to achieve a larger gravitational mass compared to its non-rotating counterpart.
In the left panel, the maximum masses for $\chi= 0.0, 0.2, 0.4, 0.6$, and $0.8$ are 2.43, 2.46, 2.52, 2.63, and 2.80$M_{\odot}$, respectively.
In particular, the maximum mass for rotating configurations $M_{\text{max}}$, can reach up to $2.93M_{\odot}$, which is 21.7\% higher than the static result $M_{\text{TOV}}$.
This result aligns with earlier studies \citep{Weber1992APJ,Cook1994APJ,Paschalidis2017LR}, which indicates that maximally rotating stars can have masses about 20\% greater than their static counterparts, underscoring the significant impact of centrifugal force on neutron star structure.

\begin{figure*}[htbp]
  \centering
  \includegraphics[width=0.45\textwidth]{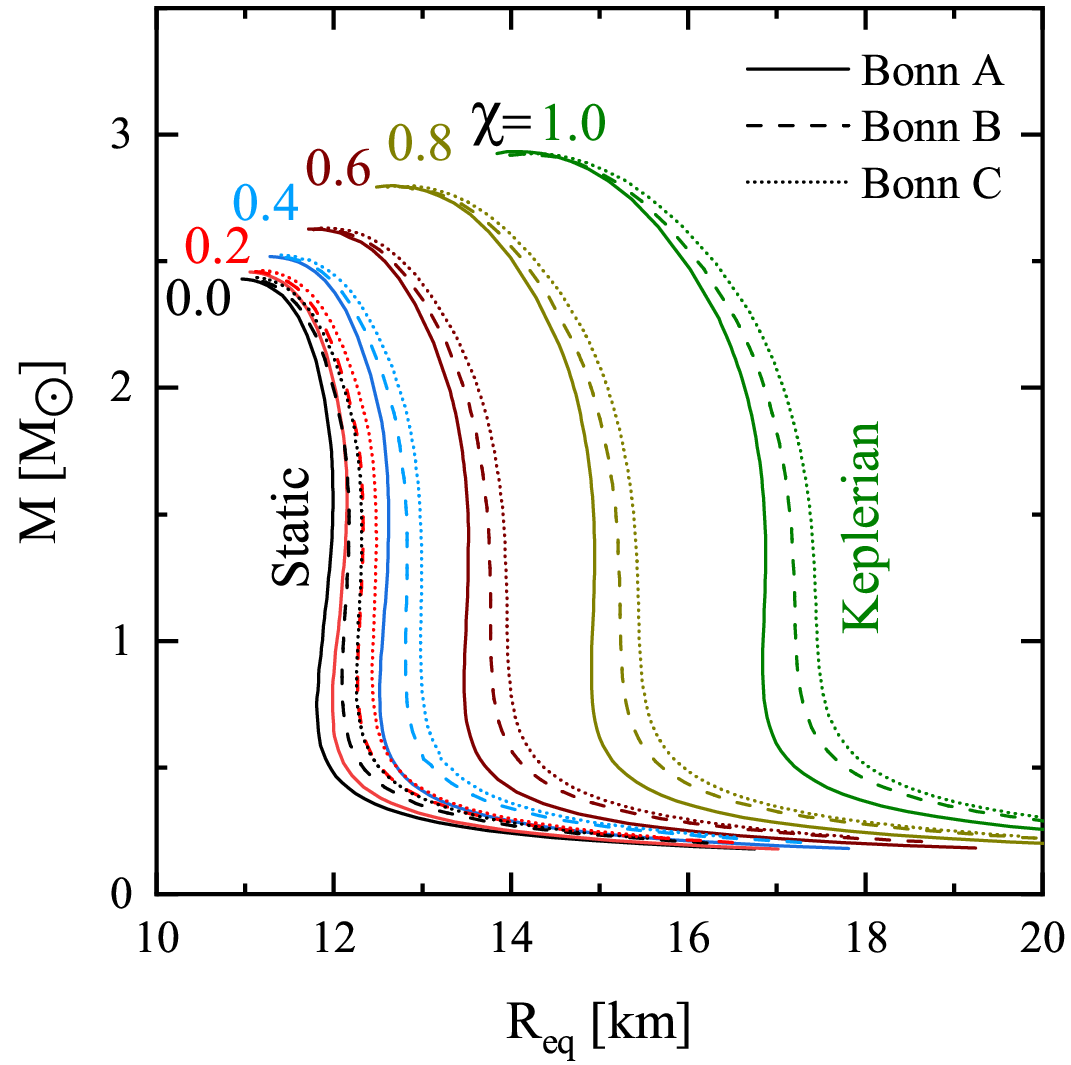}
  \hspace{0.5cm}
  \includegraphics[width=0.45\textwidth]{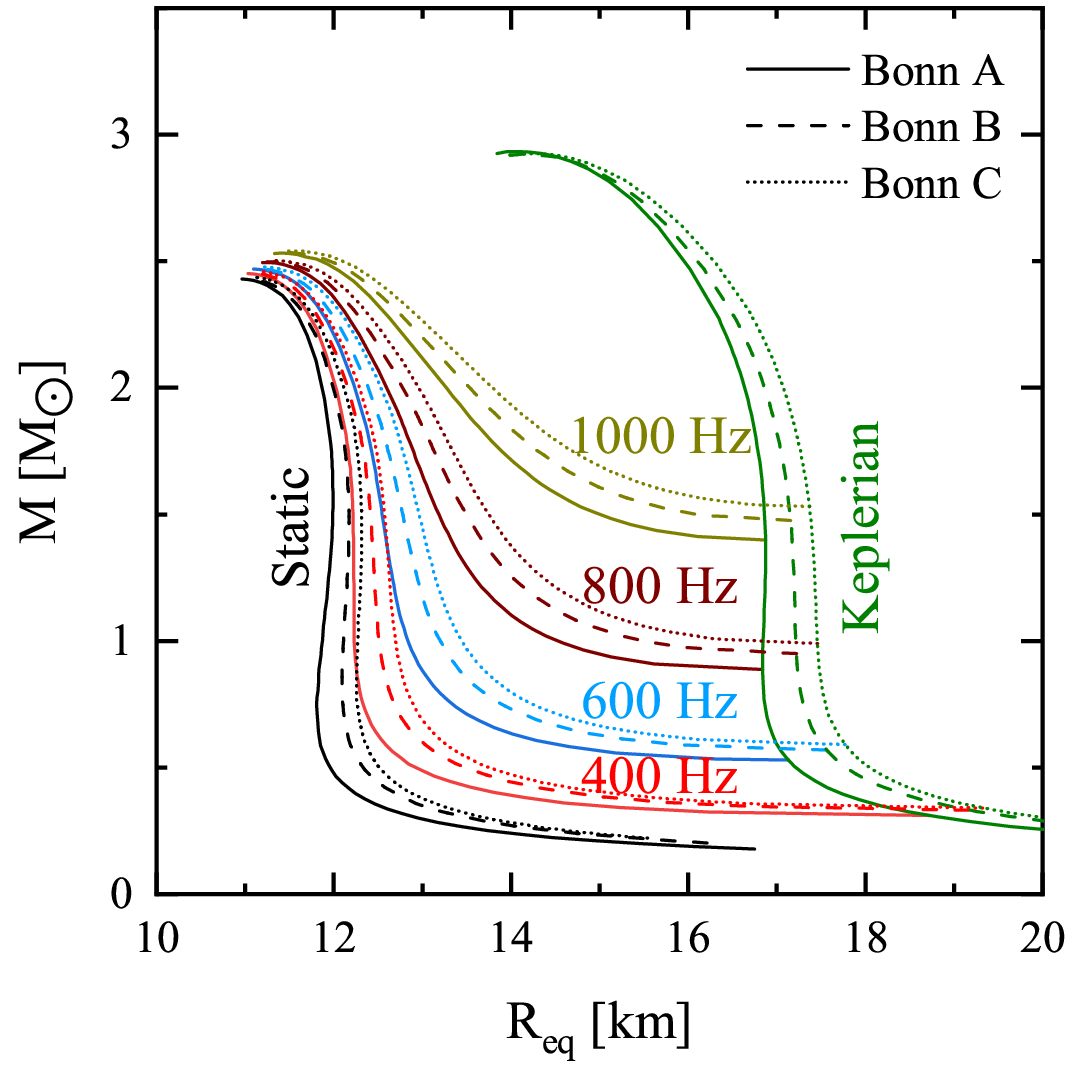}\\
  \caption{
  The gravitational mass $M$ as a function of equatorial radii $R_{\rm eq}$. Six cases are shown for (left panel) constant spin ratios  $\chi = 0$ (static), 0.2, 0.4, 0.6, 0.8, and 1.0 (Keplerian sequence) and (right panel) constant spin frequencies $f = 0, 400, 600, 800, 1000$ Hz and $f=f_K$. Results are obtained with EOSs from RBHF calculations in the full Dirac space starting from the Bonn A (solid), B (dashed), and C (dotted) potentials.}
  \label{fig2}
\end{figure*}

In Figure \ref{fig2}, the gravitational masses of both static and rotating neutron stars are plotted as functions of their equatorial radii.
Similar to Figure \ref{fig1}, the left panel represents cases with constant spin ratios $\chi$ of 0.0, 0.2, 0.4, 0.6, 0.8, 1.0, while the right panel shows cases with constant spin frequencies $f$ of 0, 400, 600, 800, 1000 Hz, and $f=f_K$.
For a fixed spin ratio or frequency, the gravitational mass decreases as the equatorial radius increases, whereas it increases with higher rotation ratios or frequencies.
For the Bonn A potential, the static case shows a  radius of 10.93 km at the maximum mass.
In contrast, at Keplerian frequency, the radius extends to 13.84 km, reflecting a 26.2\% increase due to the additional support provided by rotation, which counteracts the gravitation pull toward the star's center.

In addition to the EOS calculated with Bonn A potential, in Figure \ref{fig2} we also show the calculated results starting from Bonn B and C potentials~\citep{Brockmann1990}. 
Generally, the patterns for mass-radius relations among three parameterizations of realistic $NN$ interactions are very similar, irrelevant to static or rotational cases.
This robust consistency across different rotation frequencies suggests that the impact of rotational dynamics on the mass-radius relation may be relatively insensitive to the specific details of the underlying interactions that govern the EOS.
Specifically, the radii of a canonical neutron star with mass $1.4M_\odot$ in the static case are $R_{1.4} = 11.98$, $12.17$, and $12.32$ km for the Bonn A, B, and C potential, respectively.
The smallest radius from Bonn A implies that the RBHF calculations with this potential result in the softest EOS.
This is attributed to the weakest tensor force in the Bonn A potential, which leads to the strongest attraction between nucleons.
These differences underscore the sensitivity of neutron star properties to the underlying $NN$ interactions and emphasize the importance of accurately modeling these interactions to predict astrophysical observables.
Related discussions on the tensor-force effects in nuclear matter from realistic $NN$ interactions can be found in \cite{WangSB2024ScienceBulletin}.

\begin{figure}[htbp]
  \centering
  \includegraphics[width=0.45\textwidth]{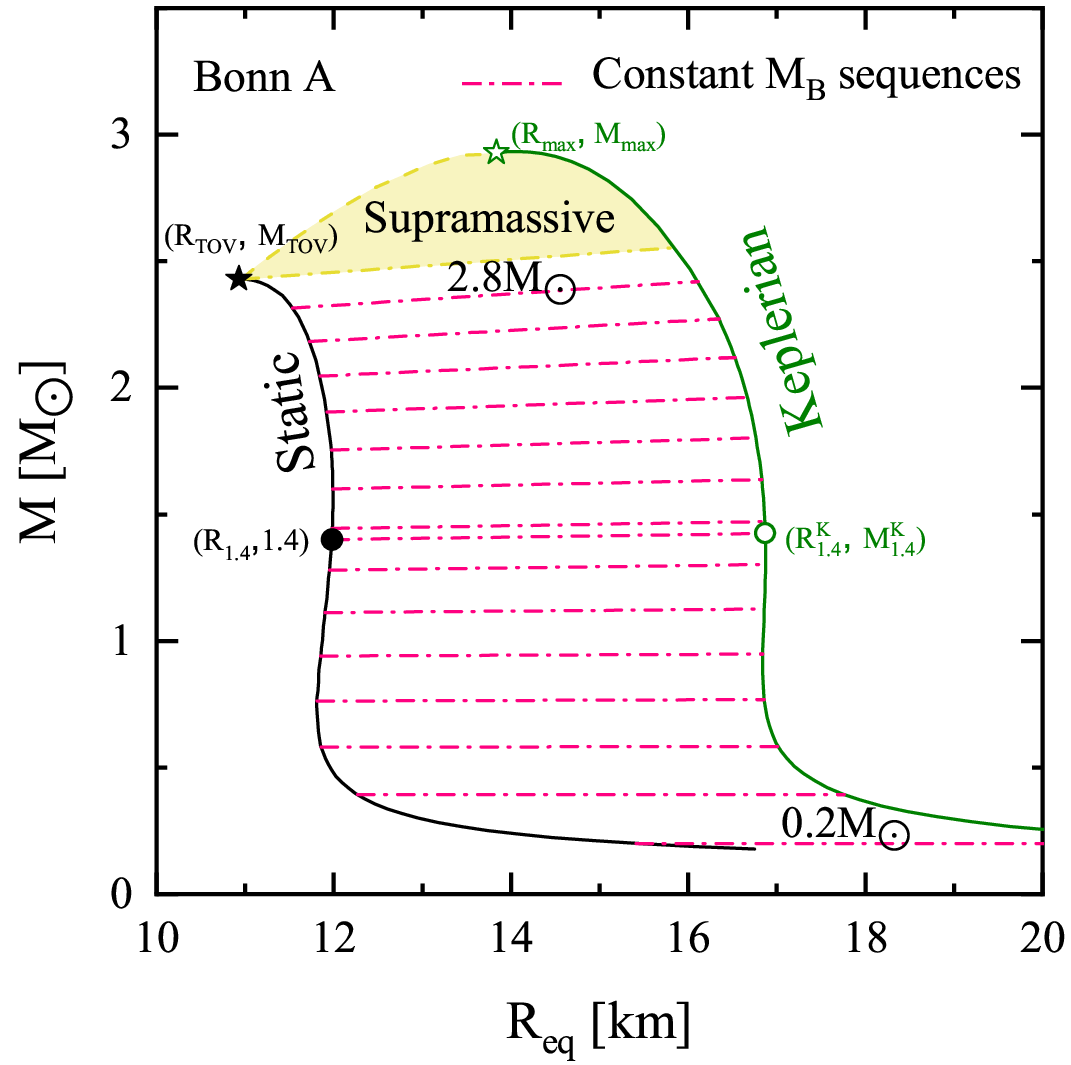}\\
  \caption{The gravitational mass $M$ as a function of equatorial radii $R_{\rm eq}$ for static neutron stars and rotating neutron ones at Keplerian limits. The constant baryonic mass sequences of $M_B = 0.2 M_{\odot} \sim 2.8 M_{\odot}$ with $0.2 M_{\odot}$ interval, are shown as roughly horizontal curves in pink. The supramassive region is depicted with yellow shadow. Selected properties of neutron stars are marked. See the text for details.}
  \label{fig3}
\end{figure}

Studying stellar sequences with constant baryonic mass $M_{B}$ provides crucial insights into the evolution of neutron stars.
Evolutionary sequences between the two extremes, i.e., non-rotating (static) stars and stars rotating at Keplerian limits, are often used to simulate the spin-down or spin-up behaviour of stars under external torques~\citep{Bildsten1998APJ,Andersson1999APJ}.
Several sequences of constant baryonic mass $M_{B}$ calculated using the RBHF theory with Bonn A potential are shown in Figure \ref{fig3}.
These sequences appear as roughly horizontal lines connecting the Keplerian sequence with the non-rotating configuration, suggesting that the gravitational mass changes barely during the spin evolution.
As the neutron star spins down along its evolutionary sequence, its radius gradually shrinks until it reaches the limiting value for a static configuration.
For a rotating neutron star without corresponding stable static configurations, known as supramassive compact star, it will eventually collapse into a black hole when the star loses its excess angular-momentum support and exceeds the maximum mass at some rotation frequency. 
The yellow shadow in Figure \ref{fig3} denotes the region of these supramassive stars, whose lower boundary is defined via the constant baryonic mass sequence corresponding to the static maximum mass.

\begin{figure}[htbp]
  \centering
  \includegraphics[width=0.45\textwidth]{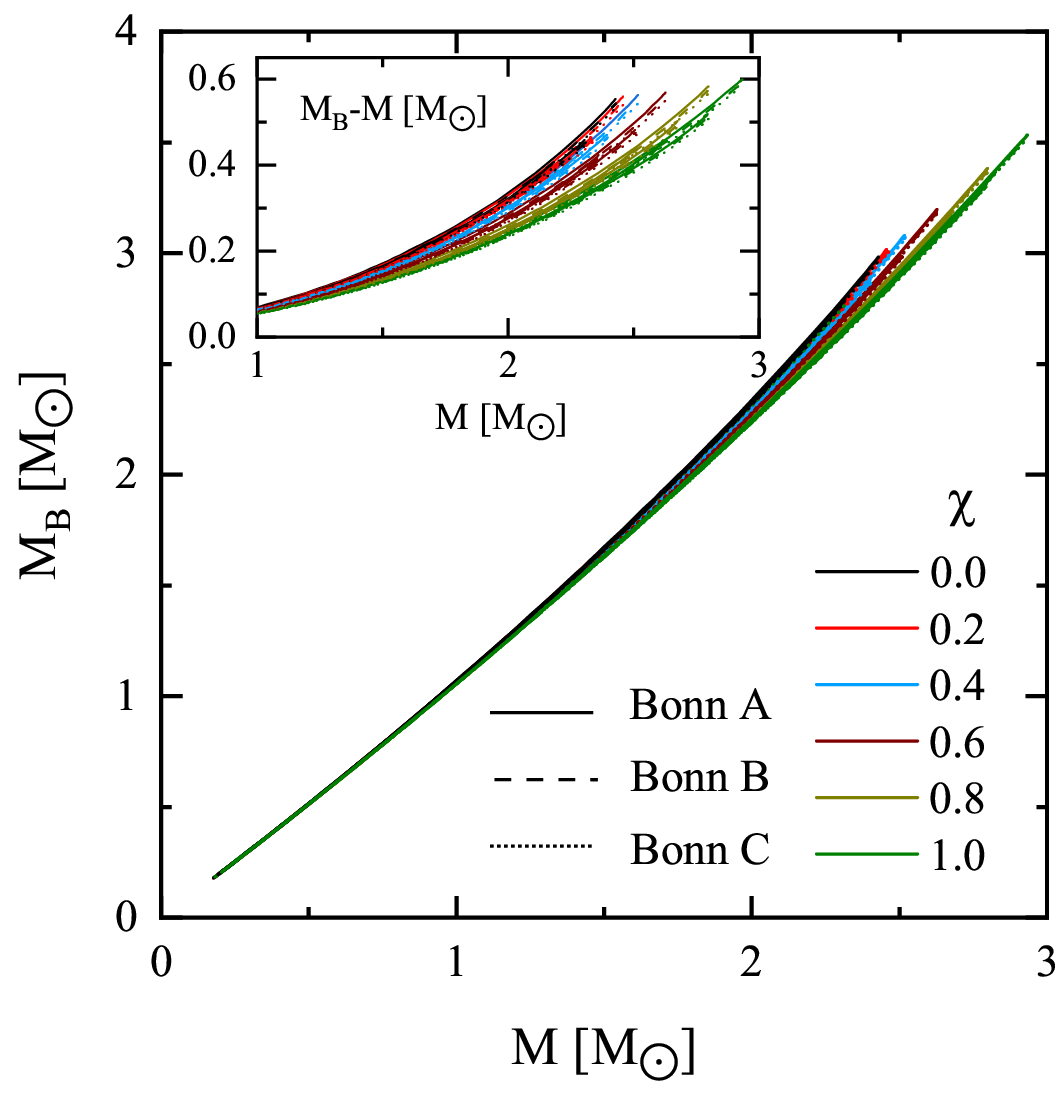}\\
  \caption{The relation between the baryonic mass $M_{B}$ and gravitational mass $M$ for a rotating neutron star with the constant spin ratio frequencies $\chi = 0$ (static), 0.2, 0.4, 0.6, 0.8, and 1.0 (Keplerian sequence) obtained by the full Dirac space, the projection method and the mom.-ind. app. with the Bonn A, B, and C potentials.
  The binding energy $M-M_B$ versus the gravitational mass for the same constant rotation frequencies is also shown in the inset.}
  \label{fig4}
\end{figure}

In addition, the conversion between the gravitational mass $M$ and baryonic mass $M_B$ is of high necessity for numerical simulations of  binary neutron star mergers \citep{2019-Baiotti-PPNP}.
The relation between the baryonic $M_B$ and the gravitational mass $M$ for a rotating neutron star with the constant spin ratio frequencies $\chi = 0$ (static), 0.2, 0.4, 0.6, 0.8, and 1.0 (Keplerian sequence) obtained by the full Dirac space, the projection method and the mom.-ind. app. method with the Bonn A, B, and C potentials are shown in Figure \ref{fig4}.
It can be seen that the baryonic mass $M_B$ increases with the gravitational mass $M$, and the values of $M_B$ obtained by different theoretical methods with different potentials at the same ratio are almost in line.
Therefore, we fit a relation between $M_B$ and $M$, incorporating $\chi$, based on the present data as follows,
\begin{equation}
  \begin{split}
  \frac{M_B}{M_{\odot}}=&\ (-0.0266\chi^2-0.0064\chi+0.1092)\left(M/M_{\odot}\right)^2\\
    &\ +(0.0219\chi^2-0.0029\chi+0.9540)(M/{M_{\odot}}).
  \end{split}
\end{equation}
The binding energy $M-M_B$ versus the gravitational mass for the same constant rotation frequencies is also shown as an inset in Figure ~\ref{fig4}.
It can be observed that the binding energies obtained by different theories and potentials are very similar at the same spin ratio. 
The differences among binding energies at different spin ratios become more pronounced as the gravitational mass increases, further demonstrating the impact of rotation on massive neutron stars.

\begin{table*}[ht]
\centering
\caption{Selected properties of static and rotating neutron stars: the maximum mass $M_{\text{TOV}}$, the corresponding radius $R_{\text{TOV}}$ and the radius $R_{1.4}$ of a $1.4M_{\odot}$ neutron star for non-rotating case, the maximum mass $M_{\text{max}}$, the corresponding equatorial radius $R_{\text{max}}$, the maximum rotation frequency $f_K$, and the eccentricity $e_K$ for Keplerian sequence.
Also shown are two quantities, $R^K_{1.4}$ and $M^K_{1.4}$, associated with a special configuration in Keplerian sequence, where its baryonic mass is the same as the one with a gravitational mass of $1.4M_{\odot}$ in static sequence.}
\begin{longtable}{ccccccccccc}
\hline \hline
Theory & Potential & $M_{\text{TOV}}$ & $R_{\text{TOV}}$ & $R_{1.4}$ & $M_{\text{max}}$ 
& $R_{\text{max}}$ & $M^K_{1.4}$      & $R_{1.4}^K$      & $f_K$     & $e_K$            \\
       &           & $[M_\odot]$      &   [km]           &   [km]    & $[M_\odot]$      
&  [km]            & $[M_\odot]$      &  [km]            &  [kHz]    &                  \\  \hline
\multirow{3}{*}{Full Dirac space} &
Bonn A  & 2.43 & 10.93 &11.98& 2.93 & 13.84 &  1.43   &16.87 & 1.86 & 0.838  \\
&Bonn B & 2.43 & 10.98 &12.17& 2.92 & 13.91 &  1.42  &17.17 & 1.84 & 0.836  \\
&Bonn C & 2.44 & 11.04 &12.32& 2.92 & 13.98 &  1.42  &17.40 & 1.83 & 0.836  \\ \hline
\multirow{3}{*}{Projection}
&Bonn A & 2.31 & 11.23 &12.37& 2.78 & 14.45 &  1.42  &17.47 & 1.71 & 0.839  \\
&Bonn B & 2.31 & 11.29 &12.53& 2.78 & 14.53 &  1.42  &17.71 & 1.69 & 0.838  \\
&Bonn C & 2.31 & 11.35 &12.65& 2.78 & 14.61 &  1.42  &17.90 & 1.68 & 0.838  \\ \hline
\multirow{3}{*}{Mom.-ind. app.}
&Bonn A & 2.17 & 11.07 &12.36& 2.61 & 14.36 &  1.42  &17.52 & 1.67 & 0.836  \\
&Bonn B & 2.18 & 11.12 &12.49& 2.61 & 14.44 &  1.42  &17.73 & 1.66 & 0.836  \\
&Bonn C & 2.17 & 11.17 &12.59& 2.61 & 14.50 &  1.42  &17.90 & 1.65 & 0.835  \\  \hline\hline
\end{longtable} \label{label-table}
\end{table*}

In Table \ref{label-table}, selected properties of both static and rotating neutron stars are shown, including the maximum mass $M_{\text{TOV}}$, its corresponding radius $R_{\text{TOV}}$, the radius of a $1.4M_{\odot}$ neutron star $R_{1.4}$ in the non-rotating case, the maximum mass $M_{\text{max}}$, its corresponding equatorial radius $R_{\text{max}}$, the maximum rotation frequency $f_K$, and the eccentricity $e_K$ (see the discussions below) for the Keplerian sequence.
In addition, we also show two quantities $R^K_{1.4}$ and $M^K_{1.4}$ associated with a special configuration in Keplerian sequence, where its baryonic mass is the same as the one with a gravitational mass of $1.4M_{\odot}$ in static sequence.
$M_{\text{TOV}}$, $R_{\text{TOV}}$, $R_{1.4}$, $M_{\text{max}}$, $R_{\text{max}}$,  $R^K_{1.4}$, and $M^K_{1.4}$ are marked in Figure \ref{fig3}.
In comparison to the results from the projection method and the mom.-ind. app., the maximum mass $M_{\text{TOV}}$ in the third column and $M_{\text{max}}$ in the sixth column for static and rotating neutron stars, derived from the full Dirac space, are the largest. 
Conversely, the corresponding radii $R_{\text{TOV}}$ and $R_{\text{max}}$ in the fourth column and the seventh column are the smallest. This indicates that the neutron star predicted by the full Dirac space is the most compact among the three approaches.
These predictions align well with current astrophysical constraints on massive neutron stars~\citep{Demorest2010Nature,Antoniadis2013,Fonseca2016APJ,NANOGrav:2017wvv, Cromartie2020NA,Fonseca2021APJ}. 
Accordingly, the radius $R_{1.4}$ of a $1.4M_{\odot}$ neutron star for non-rotating case obtained by the full Dirac space with Bonn A potential is the smallest. 
For Keplerian configurations, our calculations in the full Dirac space from the three different Bonn potentials predict the radius $R^K_{1.4}$=$16.87$, 17.17, and 17.40~km.
These values underscore the substantial influence of rapid rotation on the structure of neutron stars, extending the radius beyond the limits set by static configurations.
The maximum rotation frequency $f_K$ obtained by the full Dirac space are 1.86, 1.84, and 1.83~kHz, with the Bonn A potential yielding the largest value.
The eccentricities $e_K$ obtained by different methods with different potentials are almost the same.

\begin{figure}[htbp]
  \centering
  \includegraphics[width=0.45\textwidth]{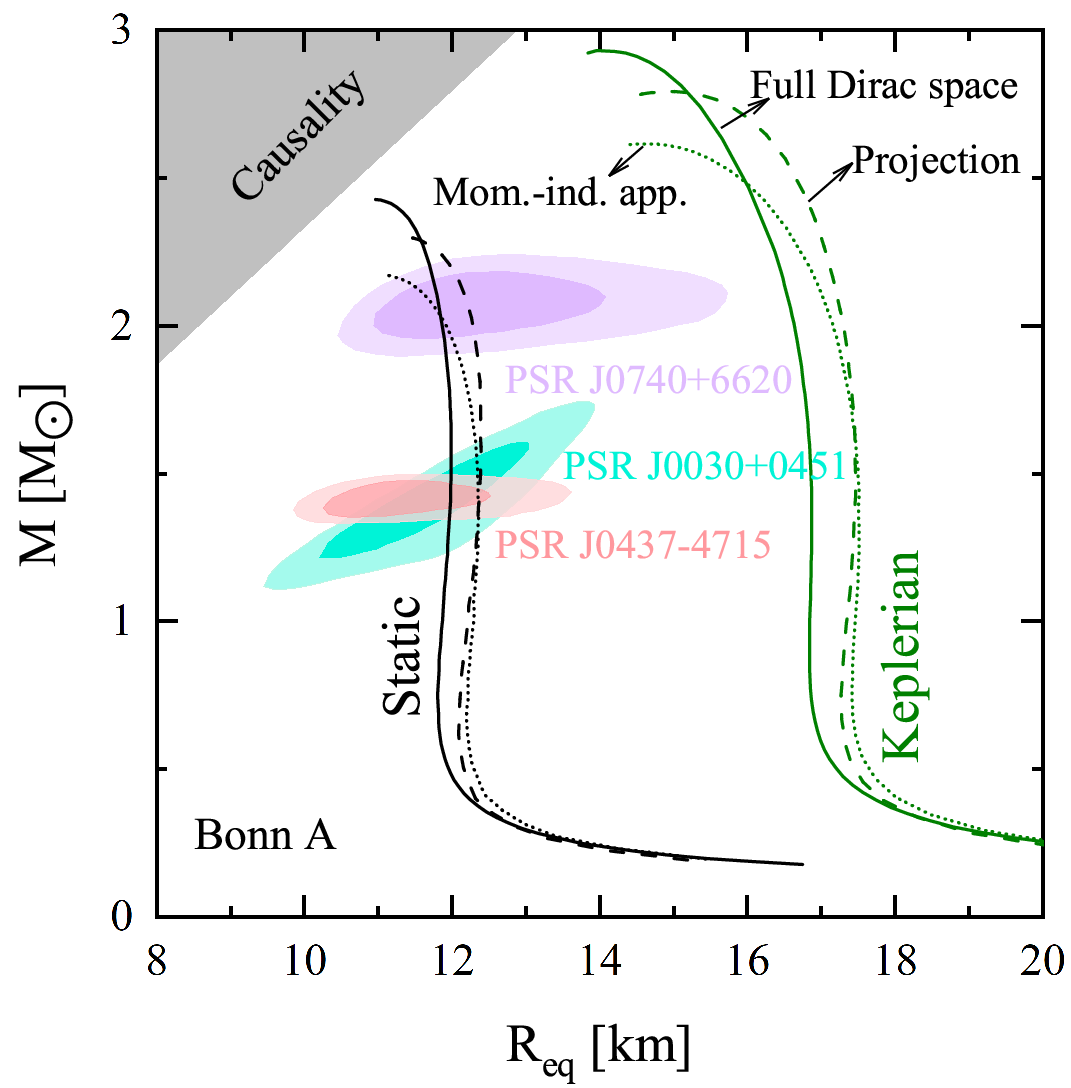}\\
  \caption{The gravitational mass $M$ as a function of equatorial radii $R_{\rm eq}$ for static and rotating neutron stars obtained by the full Dirac space (solid line), the projection method (dashed line) and  the mom.-ind. app. (dotted line) with Bonn A potential.
  The 68\% and 95\% credible regions of joint estimations on mass and radius for PSR J0437-4715 \citep{Choudhury2024APJ}, PSR J0740+6620 \citep{Salmi2024arXiv}, and PSR J0030+0451 \citep{Vinciguerra2024APJ} from NICER are also shown, with light and dark shaded regions.
  The gray shaded area in the upper left corner denotes the constraints from the causality.}
  \label{fig5}
\end{figure}

To further compare our results of mass-radius relations to astrophysical constraints, we show in Figure \ref{fig5} the gravitational mass $M$ as a function of equatorial radii $R_{\rm eq}$ for static and rotating neutron stars obtained by the full Dirac space (solid line), the projection method (dashed line) and the mom.-ind. app. (dotted line) with Bonn A potential.
Following \cite{Rutherford_2024_ApJ} 68\% and 95\% credible regions (CIs) of mass-radius for the PSR J0740+6620 results
from \cite{Salmi2024arXiv}, the PSR J0437-4715 results
from \cite{Choudhury2024APJ}, the PSR J0030+0451 results
from \cite{Vinciguerra2024APJ} are also shown in Figure \ref{fig5}.
Clearly, our results are consistent with the constraints from NICER.

Due to the absence of analytical solutions for rotating neutron stars, there appears some numerical estimations for the Keplerian frequency.
A significant number of empirical formulas for the Keplerian frequency have been produced over the years.
One of the widely used formula is given by \cite{Shapiro1989Nature}
\begin{equation}\label{fk-rm}
  f_K=\mathcal C_{st} \left( \frac{M^{st}}{M_{\odot}} \right)^{1/2}\left( \frac{10~{\rm km}}{R^{st}} \right)^{3/2},
\end{equation}
where $st$ represents for the static case.
This relation is well established, but the unknown parameter $\mathcal C_{st}$ depends highly on the various approximations and, of course, the selected EOSs.
By using the static neutron star mass, radius, and the corresponding Keplerian frequency obtained by the full Dirac space, the projection method, and the mom.-ind. app. method with Bonn A, B, and C potentials, with $M^{st}\in [0.7,2.1]M_{\odot}$, the best fit produces the value of $\mathcal C_{st} = 1.26$~kHz in our calculations.
The selection of mass intervals is based on a thorough evaluation of optimal fitting and astronomical observations.

From the equation \eqref{fk-rm}, once the mass and rotation frequency of a neutron star are known, its corresponding radius can be estimated.
The black widow PSR~J0952-0607 is the fastest known rotating pulsar in the disk of the Milky Way, with a frequency of 707~Hz and a mass of $2.35M_{\odot}$~\citep{Romani_2022_APJ}.
Using formula \eqref{fk-rm} and $f\leq f_K$, we can estimate the upper limit of the black widow pulsar's radius to be 19.85~km. 
In addition, the smallest radius at a frequency of 707~Hz and a mass of $2.35M_{\odot}$ in our calculation is $11.89$~km. Therefore, we can roughly estimate the range of the black widow pulsar's radius to be $11.89$~km$\leq R\leq19.58$~km.
The radius estimation of the PSR J1748-2446ad is not available, due to the lack of the information on its mass.

\begin{figure}[htbp]
  \centering
  \includegraphics[width=0.45\textwidth]{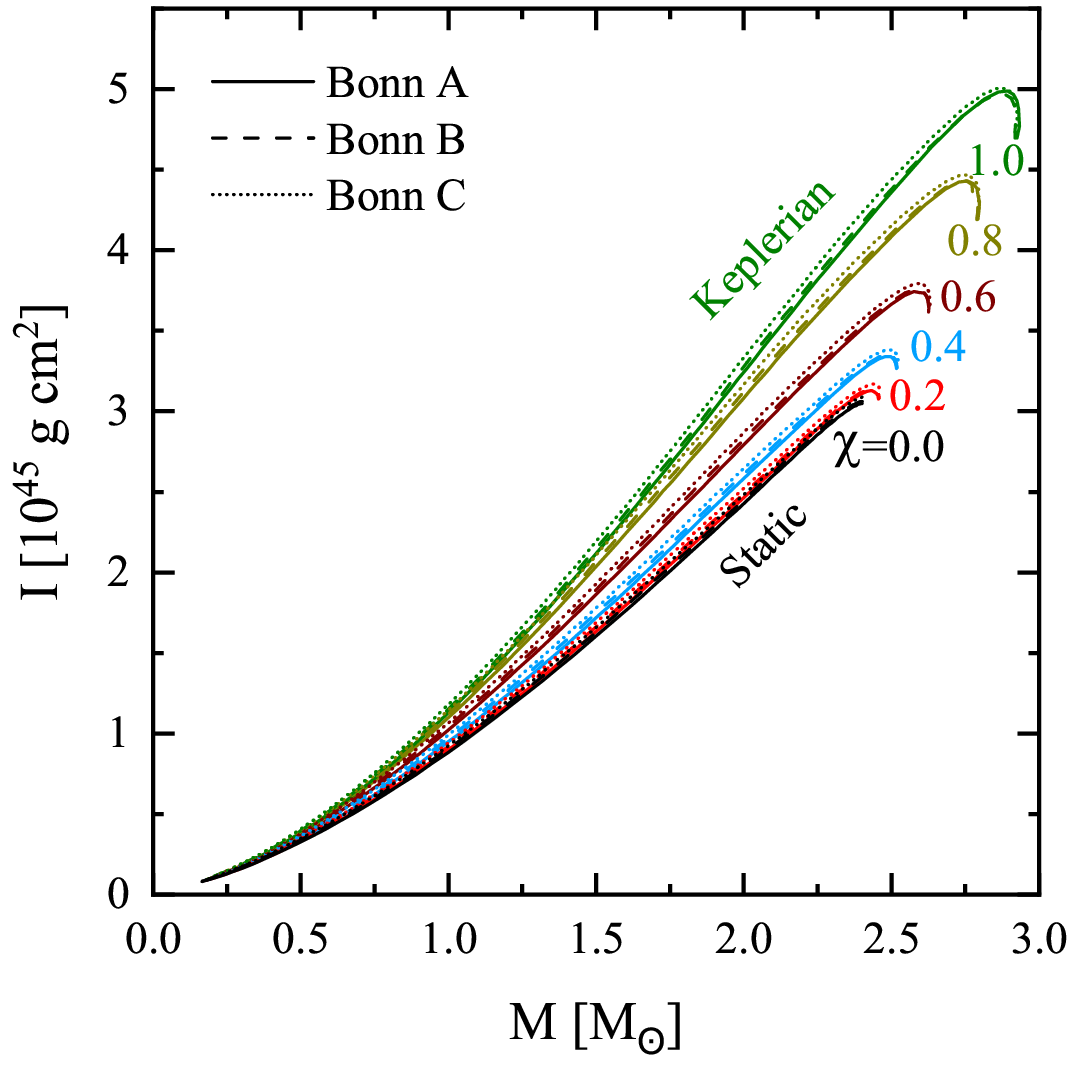}\\
  \caption{Relation between the moment of inertia and gravitational mass for several constant rotation frequencies and Keplerian frequency.}
  \label{fig6}
\end{figure}

In addition to a neutron star's mass and radius, its rotation also significantly impacts the moment of inertia, an essential parameter for pulsar analysis.
The moment of inertia $I$ quantifies how fast a neutron star can spin with a given angular momentum, and can be estimated using the relation $I = J/\Omega$, where $J$ is the star's angular momentum and $\Omega$ is the angular velocity.
Figure~\ref{fig6} displays the moment of inertia as a function of mass for various constant rotation frequencies, including the Keplerian sequence.
This figure illustrates that the moment of inertia increases with rotation frequency at a given neutron star mass.
As the star rotates more rapidly, both its radius and moment of inertia increase, reflecting the increased centrifugal forces and mass distribution further from the axis of rotation.
This enhancement in moment of inertia with higher rotational frequency has significant implications for understanding pulsar dynamics and their evolution. 
A higher moment of inertia can influence various properties of neutron stars, including their spin-down rates and the characteristics of gravitational wave signals~\citep{Abbott2021APJL}.

\begin{figure}[htbp]
  \centering
  \includegraphics[width=0.45\textwidth]{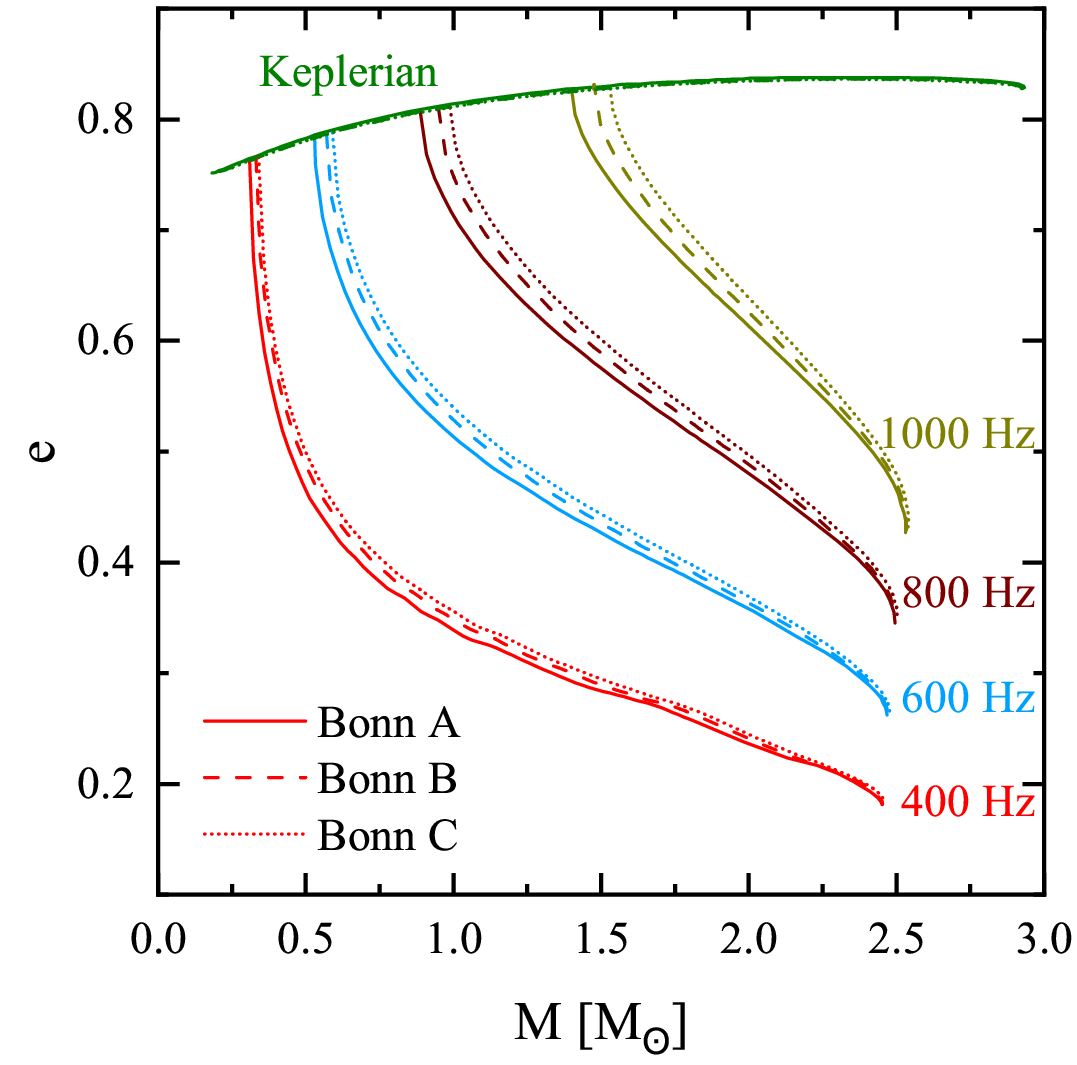}\\
  \caption{Relation between the eccentricity and gravitational mass for several constant rotation frequencies and Keplerian frequency.}
  \label{fig7}
\end{figure}

Finally, to understand how rotation deforms the equilibrium shape of a neutron star, the eccentricity is also calculated by using the formula,
\begin{equation}
	e = \sqrt{1 - \left(\frac{R_{\rm pol}}{R_{\rm eq}}\right)^2},
\end{equation}
where $ R_{\rm pol} $ and $ R_{\rm eq} $ represent the polar and equatorial radii.
In Figure \ref{fig7}, we plot the eccentricity as a function of the gravitational mass $M$ of rotating neutron stars with several constant rotation frequencies and Keplerian frequency.
For a given rotation frequency, the eccentricity decreases with increasing gravitational mass, indicating that more massive neutron stars are less deformed by rotation because their stronger gravitational pull, which counteracts the centrifugal forces.
For a fixed mass, the eccentricity increases with increasing the rotation frequency.
As the star spins more rapidly, stronger centrifugal forces lead to increased polar flattening and equatorial bulging, thereby increasing the eccentricity.
One noticeable point is the consistency among three realistic $NN$ interactions for the Keplerian sequences, implying that the eccentricity for the Keplerian frequency is a universal property largely independent of the specific EOS.
Therefore, the relation among eccentricity, gravitational mass, and rotational frequency highlights the complex interplay between gravitational and centrifugal forces in shaping the structure of rotating neutron stars.
Understanding this deformation is crucial for accurately describing neutron star properties and interpreting astrophysical observational data.

\begin{figure}[htbp]
  \centering
  \includegraphics[width=0.45\textwidth]{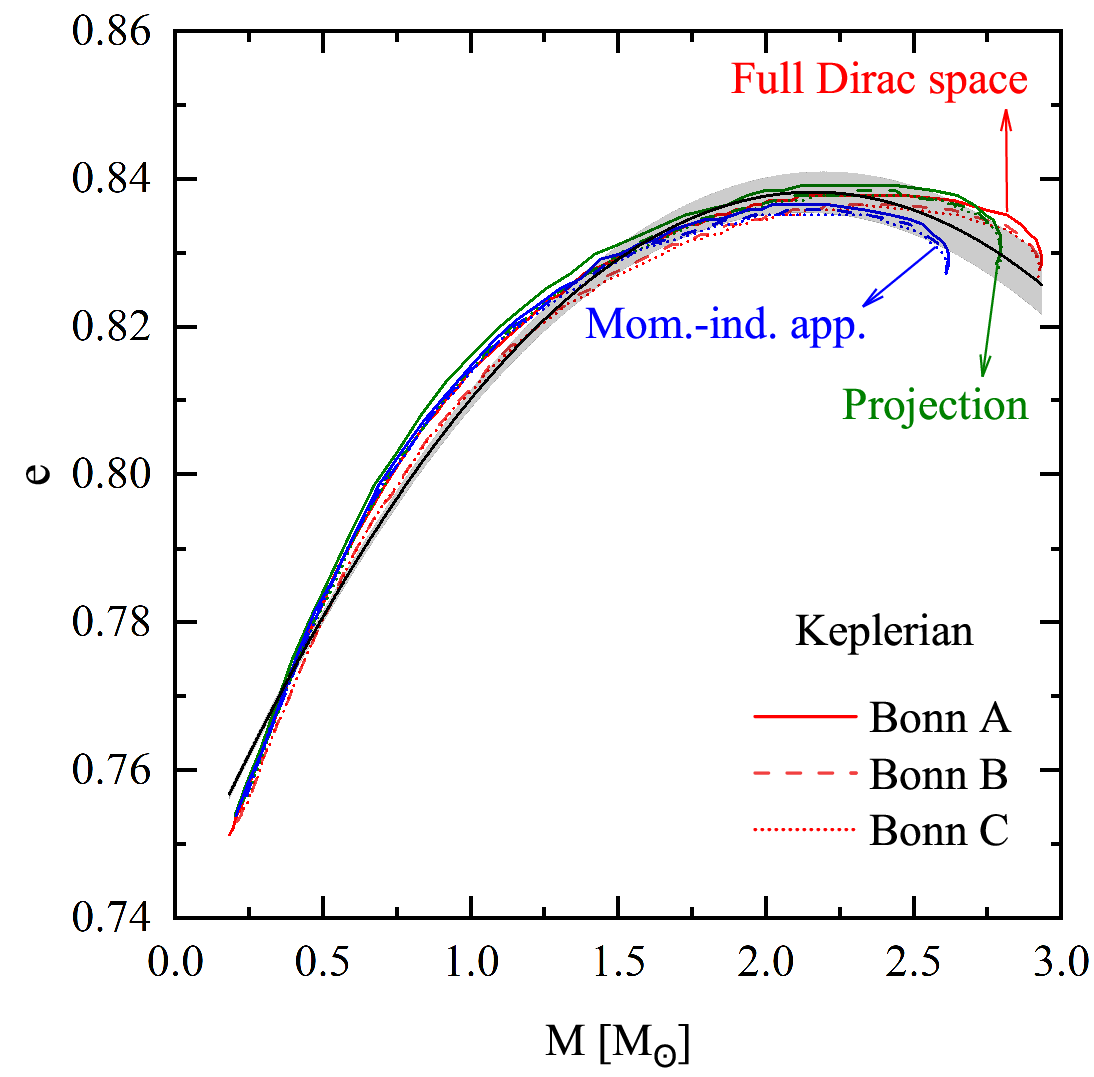}\\
  \caption{Relation between the eccentricity at the Keplerian frequency and gravitational mass obtained by using the full Dirac space method, the projection method, and the mom.-ind. app. method with Bonn A, B, and C potentials. The solid black line and the gray shaded area are the central value and errors obtained by fitting our results.}
  \label{fig8}
\end{figure}

Figure \ref{fig8} shows the relation between eccentricity at the Keplerian frequency and gravitational mass, obtained by using the full Dirac space method, the projection method and the mom.-ind. app. method with Bonn A, B, and C potentials.
The eccentricities obtained by the same method with different potentials show close agreement, while values obtained by different methods exhibit more noticeable variation.
However, these deviations remain very small due to the relatively fine scale of the ordinate.
To provide a more precise description, we fit the relation between eccentricity $e$ at the Keplerian frequency and gravitational mass $M$ as follows,
\begin{equation}
  e=-0.0335\left(\frac{M}{M_{\odot}}\right)^2+0.1188\left(\frac{M}{M_{\odot}}\right)+0.7358.
\end{equation}
The shaded area represents the error range for this fit.

\section{Summary}\label{summary}

The structural properties of rotating neutron stars are studied with equation of states (EOSs) calculated by the relativistic Brueckner-Hartree-Fock theory in the full Dirac space, with Bonn potentials as realistic nucleon-nucleon interactions.
For a given central density, the inclusion of rotation enables a neutron star to achieve a gravitational mass significantly higher than the non-rotating counterpart.
For Bonn A potential, the maximum mass for rotating configurations and its corresponding radius can reach up to $2.93M_{\odot}$ and 13.84 km, which are 21.7\% and 26.2\% higher than the static results, respectively, reflecting significant impact of centrifugal force that pushes the limits of mass and radius beyond those of static configurations.
Meanwhile, similar patterns for mass-radius relations among different realistic nucleon-nucleon interactions and rotation frequencies are obtained, suggesting that the impact of rotational dynamics on the mass-radius relation may be relatively insensitive to the specific details of the underlying interactions that govern the EOS.
By confronting the calculated mass-radius relations of both the static and fast rotating neutron stars to recent astronomical observations of massive neutron stars and simultaneous mass-radius measurements, our \textit{ab initio} predictions are harmony with these observation constraints.
By using the approximately valid universal relations between the Keplerian frequency and the static mass as well as radius, the black widow PSR J0952-0607 is predicted to be less than 19.58 km.

Several sequences of constant baryonic mass $M_{B}$ are obtained. These sequences appear as roughly horizontal lines connecting the Keplerian sequence with the non-rotating configuration, suggesting that the gravitational mass changes barely during the spin evolution. 
A relation between baryonic mass $M_B$ and gravitational mass $M$ with different spin ratios with respect to Keplerian frequency are found. 
This also provides important information for the binding energies which might be related to violent neutron star mergers.

In order to better understand the rapidly rotating neutron star, the evolution of the moment of inertia and the eccentricity with the gravitational mass are also studied. Both the moment of inertia and eccentricity increases with rotation frequency. 
For the letter, one notices that there is a universal consistency between the eccentricity at the Keplerian frequency and the gravitational mass and a relation is obtained with best fit. 
This would enhance our understandings for the properties of rotating neutron stars especially at the Keplerian frequency.

\section{Acknowledgment}
\begin{acknowledgments}
This work was partly supported by the Guizhou Provincial Science and Technology Projects under Grant No. ZK[2022]203, PhD fund of Guizhou Minzu University under Grant No. GZMUZK[2024]QD76, the National Natural Science Foundation of China under Grant Nos. 12205030, 12265012, and the Project No. 2024CDJXY022 supported by the Fundamental Research Funds for the Central Universities. Part of this work was achieved by using the supercomputer OCTOPUS at the Cybermedia Center, Osaka University under the support of Research Center for Nuclear Physics of Osaka University.
Part by the European Research Council (ERC) under the European Union's Horizon 2020 research and innovation programme (AdG EXOTIC, grant agreement No. 101018170) and by the MKW NRW under the funding code NW21-024-A.
\end{acknowledgments}


\bibliography{reference}{}
\bibliographystyle{aasjournal}

\end{CJK*}
\end{document}